\documentstyle[12pt]{article}
\def\a{\alpha}
\def\b{\beta}

\def\d{\delta}
\def\e{\epsilon}

\def\g{\gamma}

\def\p{\psi}

\def\l{\lambda}
\def\m{\mu}

\def\o{\omega}

\def\r{\rho}

\def\vf{\varphi}

\def\be{\begin{equation}}
\def\ee{\end{equation}}
\def\arr{\begin{array}{rll}}
\def\ea{\end{array}}
\def\bea{\begin{eqnarray}}
\def\eea{\end{eqnarray}}

\begin{document}

\begin{titlepage}
\noindent

\vskip 1.5cm

\begin{center}

{\Large\bf Can one restore Lorentz invariance }\\

\vskip 0.5cm

{\Large\bf in quantum N=2 string? }\\

\bigskip

\vskip 1.5cm

{\large Stefano Bellucci}~\footnote{
E-mail: Stefano.Bellucci@lnf.infn.it} 
and 
{\large Anton Galajinsky}~\footnote{
On leave from Department of Mathematical Physics,
Tomsk Polytechnical University, Tomsk, Russia\\
\phantom{XX}   
E-mail: Anton.Galajinsky@lnf.infn.it}

\vskip 0.4cm

{\it  INFN--Laboratori Nazionali di Frascati, C.P. 13, 
00044 Frascati, Italy}
\vskip 0.4cm

\end{center}

\vskip 1.5cm

\begin{abstract}
\noindent
We consider quantum $N=2$ string embedded into the $N=4$ topological
framework from the perspective of the old covariant quantisation. 
Making use of the causality and cyclic symmetry
of tree amplitudes we argue that no Lorentz covariant boson emission
vertex can be constructed within the $N=4$ topological formalism.
\end{abstract}

\vspace{0.5cm}


\end{titlepage}

\noindent
{\bf 1. Introduction}\\[-4pt]

\vskip 0.4cm

Critical $N=2$ string provides a conventional framework for describing
self--dual gauge theory or self--dual gravity in two spatial and two
temporal dimensions. Apart from a number of salient features
characterising the model like the absence of massive excitations
in the spectrum of physical states,
continuous family of sectors interpolating between
$R$ and $NS$ and connected by spectral flow,
the vanishing of scattering amplitudes with more than three external
legs to all orders in perturbation theory, there is a fundamental
drawback intrinsic to the model. It lacks manifest Lorentz covariance.
Technically, in order to construct the $U(1)$ current entering  
the $N=2$ superconformal algebra (SCA) out of the fermionic fields at hand
one has to introduce a complex structure in the target which breaks 
the full Lorentz group $SO(2,2)$ down to the $U(1,1)$ subgroup.

A way out of the problem has been observed by Siegel~\cite{ws}
and elaborated in much more detail by Berkovits and Vafa~\cite{berkvafa}
(see also the related works~\cite{ov3,bvw}). The idea is that one
can embed the $N=2$ string in a larger $N=4$ topological framework
by adding to the theory two more fermionic currents and two more
bosonic ones which extend the $N=2$ SCA to a small $N=4$ SCA. Classically
the new constraints prove to be functionally dependent on those
forming the $N=2$ SCA~\cite{ws,bg}. The crucial observation, however,
is that extending the $N=2$ algebra to a small $N=4$ algebra one 
raises also the group of external global automorphisms 
($U(1,1)$ in the case on the $N=2$ string) to the full Lorentz group.
In other words, 
within the $N=4$ topological framework one reveals a tempting possibility 
to restore the Lorentz invariance missing in the $N=2$ formalism. 
Curiously enough, this resembles the Green--Schwarz superstring,
for which the extracting of a functionally independent set of 
fermionic first class constraints is known to break the manifest Lorentz 
covariance. 
 
A classical action for the $N=4$ topological extension of the $N=2$
string has been constructed recently in Refs.~\cite{bg,bg1}.
In order to provide the higher global symmetry, on the world--sheet 
of the string some extra fields are to be introduced, these 
complementing the $d=2$, $N=2$ conformal supergravity 
multiplet to $d=2$, $N=4$ one. The global limit of the four local 
supersymmetry transformations corresponds to a twisted version
of the $N=4$ supersymmetry algebra~\cite{bdg}. Interestingly enough, 
being classically 
equivalent on a flat background, the theories lead to different 
geometries when put in a curved space. In contrast to the K\"ahler geometry
characterising the $N=2$ case, for the $N=4$ model a manifold
has to admit a covariantly constant holomorphic two--form
in order to support the $N=4$ twisted supersymmetry~\cite{bdg}. This restricts
the holonomy group to be a subgroup of $SU(1,1)$ and leads to a 
Ricci--flat manifold already at the classical level.

Turning to the quantum description for the $N=4$ model, 
the reducibility of the constraints causes a serious complication of
the BRST procedure~\footnote{In contrast to the Green--Schwarz superstring,
the constraints intrinsic to the $N=4$ model are scalars. Hence
they do not seem to require infinitely
many ghost fields.} and a rigorous BRST quantisation of the $N=4$ topological
string is unknown. To bypass the problem, in Ref.~\cite{berkvafa}
(see also~\cite{ov3,bvw}) a set of reasonable prescriptions to calculate 
scattering amplitudes has been proposed and shown to reproduce
the results known for the $N=2$ string
including an elegant proof of the vanishing theorems.
Notice, however, that the prescription essentially relies upon a specific 
topological twist which does not treat all the currents on equal 
footing and breaks the Lorentz group down to the $U(1,1)$ subgroup.

The purpose of this paper is to reconsider the issue of the Lorentz 
invariance in the quantum $N=4$  topological string
from the perspective of the old covariant quantisation. Avoiding 
problematic BRST analysis this still maintains one within a 
conventional framework. In the next 
section we briefly discuss the extension of the $N=2$ SCA to a small $N=4$
SCA and give an explicit form of the Lorentz transformations arising
within the $N=4$ topological framework. Sect. 3 contains the discussion of 
vertex operators. In particular, we show that, although the vertex 
operators describing asymptotic physical states do respect the full Lorentz
group, the causality and cyclic symmetry of tree amplitudes
prevent one from constructing a Lorentz invariant boson emission 
vertex. Thus, the conclusion we draw is that, 
although giving an efficient key to the restoration
of the Lorentz invariance at the classical level, the $N=4$ formalism
fails to do so at the quantum level.
 
\vskip 0.5cm

\noindent
{\bf 2. N=2 SCA, small N=4 SCA and Lorentz symmetry}\\[-4pt]

\noindent
\vskip 0.4cm

Let us consider a $c=6$ CFT corresponding to the critical closed $N=2$ string.
The matter sector involves two complex bosons $x^a(z,\bar z)$, $a=0,1$,
and four complex fermions $\p^a (z)$, $\vf^a (\bar z)$,
the latter belonging to the right and left movers, respectively.
The fields are assigned with the standard 
propagators\footnote{On the cylinder we conjugate as
${(x^a)}^{*}={\bar x}^{\bar a}$, ${(\p^a)}^{*}={\bar \p}^{\bar a}$.
The target metric is Hermitian ${\eta^{\bar a b}}^{*}=\eta^{\bar b a}$.} 
\bea
&&
\langle x^a(z,\bar z)~ {\bar x}^{\bar a}(z',\bar z') \rangle=
\langle {\bar x}^{\bar a}(z,\bar z)~ x^a(z',\bar z') \rangle=
-\eta^{\bar a a} (\ln (z-z')+\ln (\bar z -\bar z')),
\nonumber\\[2pt]
&&
\langle \p^a(z)~ {\bar\p}^{\bar a}(z') \rangle=
\langle {\bar\p}^{\bar a}(z)~ \p^a(z') \rangle=
-\frac {\eta^{\bar a a}}{z-z'},
\nonumber\\[2pt]
&&
\langle \vf^a(\bar z)~ {\bar\vf}^{\bar a}(\bar z') \rangle=
\langle {\bar\vf}^{\bar a}(\bar z)~ \vf^a(\bar z') \rangle=
-\frac {\eta^{\bar a a}}{\bar z-\bar z'},
\eea
where $\eta^{\bar a a}=\mbox{diag}~(-,+)$ stands for the
Minkowski metric in the target.

Given the matter fields, one can readily construct the $N=2$ 
superconformal currents (in what follows we discuss only the right 
movers and use the abbriviation $\vf \bar\chi=\vf^a \eta_{a\bar a}
{\bar\chi}^{\bar a}$)
\bea
&&
T(z)=-\partial x \partial \bar x +{\textstyle{\frac {1}{2}}}
(\p \partial \bar\p+\bar\p \partial \p),
\nonumber\\[2pt]
&&
G(z)=\partial x \bar\p, \qquad \bar G (z)=\partial\bar x \p,
\nonumber\\[2pt]
&&
J(z)=\bar\p \p.
\eea
The corresponding OPE's are well known
\bea
&&
T(z)~T(z')\sim \frac{3} {{(z-z')}^4} +\frac{2 T(z')}{{(z-z')}^2}
+\frac{\partial T(z')}{z-z'},
\nonumber\\[2pt]
&&
T(z)~G(z')\sim \frac{3}{2} \frac{G(z')}{{(z-z')}^2}
+\frac{\partial G(z')}{z-z'},
\nonumber\\[2pt]
&&
T(z)~\bar G (z')\sim \frac{3}{2}\frac{\bar G (z')}{{(z-z')}^2}
+\frac{\partial \bar G (z')}{z-z'},
\nonumber\\[2pt]
&&
T(z)~J(z')\sim \frac{J(z')}{{(z-z')}^2}
+\frac{\partial J(z')}{z-z'},
\nonumber\\[2pt]
&&
G(z)~\bar G (z')\sim \frac{2} {{(z-z')}^3} -\frac{J(z')}{{(z-z')}^2}
-\frac{1}{2}\frac{\partial J(z')}{z-z'} +\frac{T(z')}{z-z'},
\nonumber\\[2pt]
&&
G(z)~J(z')\sim \frac{G(z')}{z-z'}, \quad
\bar G(z)~J(z')\sim -\frac{\bar G(z')}{z-z'},
\nonumber\\[2pt]
&&
J(z)~J(z')\sim\frac{2}{{(z-z')}^2},
\eea
which also imply that
the fermionic currents $G$ and $\bar G$ carry confromal spin $3/2$
while the bosonic $U(1)$ current $J$ has conformal spin $1$.   
Notice that with respect to the latter the generators 
$G(z)$ and $\bar G(z)$ are charged with the charges $-1$ and $+1$, 
respectively.

A striking point about the $N=2$ algebra is that it admits a continuous
automorphism~\cite{schwimmer} with a local parameter $\a(z)$
\bea\label{sflow}
&&
T'=T-i\partial\a J+{(i\partial\a)}^2,
\quad J'=J-2i\partial\a,
\nonumber\\[2pt]
&&
G'=e^{i\a}G,\qquad \bar G'=e^{-i\a} \bar G,
\eea
which relates R and NS sectors (spectral 
flow) and allows one to stick with a preferred representation.
For the rest of the paper we choose to work in the NS representation.
Since, by the very construction, each current in the $N=2$ SCA holds 
invariant under the action of the $U(1,1)$ group, the latter provides
a global automorphism of the $N=2$ algebra which also coincides with the
global symmetry group intrinsic to the $N=2$ string.

With a closer inspection one can further discover that
some extra currents can be constructed
out of the matter fields at hand~\cite{ws,berkvafa}
\bea\label{top1}
&&
H(z)=\underline{\partial x \p}, \qquad \bar H(z)=\underline{\partial 
{\bar x} {\bar\p}},
\nonumber\\[2pt]
&&
J^{(1)}(z)=\underline{\p \p}, \qquad J^{(2)}=\underline{{\bar\p}
{\bar\p}},
\eea
where we denoted $\underline{\vf\p}=\vf^a \e_{ab} \p^b$,  
$\underline{{\bar\vf}{\bar\p}}={\bar\vf}^{\bar a} \e_{\bar a \bar b} 
{\bar\p}^{\bar b}$ and $\e_{ab}={(\e_{\bar a \bar b})}^{*}$, 
$\e_{01}=-1$, is the Levi-Civita totally antisymmetric tensor.
These extend the $N=2$ SCA to a small $N=4$ SCA
\bea
&&
T(z)~H(z')\sim \frac{3}{2} \frac{H(z')}{{(z-z')}^2}
+\frac{\partial H(z')}{z-z'},
\nonumber\\[2pt]
&&
T(z)~\bar H (z')\sim \frac{3}{2}\frac{\bar H (z')}{{(z-z')}^2}
+\frac{\partial \bar H (z')}{z-z'},
\nonumber\\[2pt]
&&
T(z)~J^{(1,2)}(z')\sim \frac{J^{(1,2)}(z')}{{(z-z')}^2}
+\frac{\partial J^{(1,2)}(z')}{z-z'},
\nonumber\\[2pt]
&&
G(z)~\bar H(z')\sim -\frac{J^{(2)}(z')}{{(z-z')}^2}-\frac{1}{2}
\frac{\partial J^{(2)} (z')}{z-z'},
\nonumber\\[2pt]
&&
\bar G(z)~H(z')\sim-\frac{J^{(1)}(z')}{{(z-z')}^2}-\frac{1}{2}
\frac{\partial J^{(1)} (z')}{z-z'},
\nonumber\\[2pt]
&&
G(z)~J^{(1)}(z')\sim-\frac{2H(z')}{z-z'}, \qquad
\bar G(z)~J^{(2)}(z')=-\frac{2\bar H(z')}{z-z'},
\nonumber\\[2pt]
&&
J(z)~H(z')\sim\frac{H(z')}{z-z'}, \qquad J(z)~\bar H(z')\sim
-\frac{\bar H(z')}{z-z'},
\nonumber\\[2pt]
&&
J(z)~J^{(1)}(z')\sim \frac{2 J^{(1)}(z')}{z-z'}, \qquad
J(z)~J^{(2)}(z')\sim -\frac{2 J^{(2)}(z')}{z-z'},
\nonumber\\[2pt]
&&
H(z)~\bar H (z')\sim -\frac{2} {{(z-z')}^3} -\frac{J(z')}{{(z-z')}^2}
-\frac{1}{2}\frac{\partial J(z')}{z-z'} -\frac{T(z')}{z-z'},
\nonumber\\[2pt]
&&
H(z)~J^{(2)}(z')\sim -\frac{2G(z')}{z-z'}, \quad
\bar H(z)~J^{(1)}(z')\sim -\frac{2\bar G(z')}{z-z'},
\nonumber\\[2pt]
&&
J^{(1)}(z)~J^{(2)}(z')\sim\frac{4}{{(z-z')}^2}+\frac{4J(z')}{z-z'}.
\eea
In checking the OPE's the identities
\be\label{idd}
\e_{ab}\eta^{\bar b b}\e_{\bar b \bar a}=\eta_{a\bar a},
\quad \eta^{\bar a a} \eta^{\bar b b} \e_{\bar a \bar b}=\e^{ab},
\ee
prove to be helpful.

Thus, altogether there are four fermionic currents of conformal spin
$3/2$. The bosonic triplet $J,J^{(1)},J^{(2)}$ (spin 1) forms an $su(1,1)$
Kac--Moody subalgebra. It is noteworthy that viewed as constraints
at the classical level the extra currents prove to be functionally 
dependent on those forming the $N=2$ algebra~\cite{ws,bg}. A remarkable
fact, however, is that extending the algebra that way, one raises the global 
automorphism group to the full Lorentz group $SO(2,2)$ and restores 
the Lorentz invariance in the classical $N=2$ string~\cite{bg,bg1}. 

Let us dwell on the issue for the case of the quantum $N=4$ algebra.
First of all, it is straightforward to verify that
the continuous automorphism~(\ref{sflow}) remains to hold in the extended
version, provided the new currents transform in accord with
\bea
&&
H'=e^{-i\a}H,\quad \bar H'=e^{i\a} \bar H, \quad
J^{(1)'}=e^{-2i\a} J^{(1)}, \quad J^{(2)'}=e^{2i\a} J^{(2)}.
\eea
Hence, one can continue to work in a preferred (NS) representation.
Turning to global automorphisms, as transformations from 
the conventional $SU(1,1)$ group 
leave each current of the $N=4$
algebra invariant, these provide an apparent automorphism. 
A less obvious point is that within the extended framework the $U(1)$ 
automorphism one had in the $N=2$ case is accompanied
by two extra transformations, altogether forming another 
${SU(1,1)}_{outer}$ group (we stick with the terminology of Ref.~\cite{bvw}).
For the elementary field 
combinations $\psi{\bar\vf}$, 
$\underline{\psi \vf}$  $\underline{{\bar\psi}{\bar\vf}}$ these read 
(independently of the statistics of the fields involved)
\vspace{0.3cm}
$$
\begin{array}{lll}
\begin{tabular}{|l|c|c|c|c|c|c|c|c|c|c|c|c|}
\hline          \vphantom{$\displaystyle\int$}
${SU(1,1)}_{outer}$  & $\d_{\o}$ & $\d_{\b}$ & $\d_{\l}$ \\
\hline          \vphantom{$\displaystyle\int$}
$\p {\bar\vf}$ & $\o(\underline{\p\vf}-\underline{{\bar\p}{\bar\vf}})$ 
& 0 & $i\l(\underline{\p\vf}+\underline{{\bar\p}{\bar\vf}})$\\
\hline          \vphantom{$\displaystyle\int$}
$\underline{\p\vf}$ & $\o(\p{\bar\vf}-{\bar\p}\vf)$ & $-2i\b 
(\underline{\p\vf})$  & 
$-i\l(\p\bar\vf -{\bar\p}\vf)$\\
\hline          \vphantom{$\displaystyle\int$}
$\underline{{\bar\p}{\bar\vf}}$ & $-\o(\p{\bar\vf}-{\bar\p}\vf)$ &
 2i$\b(\underline{{\bar\p}{\bar\vf}})$  & 
$-i\l(\p\bar\vf -{\bar\p}\vf)$\\
\hline
\end{tabular}
\end{array}
$$
\vspace{0.5cm}
where $\o$, $\b$ and $\l$ are real infinitesimal constant parameters.
We have put the $U(1)$ automorphism mentioned above in the 
second column. Being applied to the $N=4$ superconformal currents the
${SU(1,1)}_{outer}$ transformations ammount to 
(we give them in an infinitesimal form)
\bea\label{automor1}
&&
T'=T, \quad G'=G-\o \bar H +\o H, \quad \bar G'=\bar G+\o \bar H -\o H,
\nonumber\\[2pt]
&&
H'=H-\o \bar G +\o G, \quad \bar H'=\bar H+\o \bar G -\o G,
\quad
J'=J+\o J^{(2)} -\o J^{(1)},
\nonumber\\[2pt]
&&
J^{(1)'}=J^{(1)}-2\o J, \quad J^{(2)'}=J^{(2)}+2\o J,
\eea
\bea
&&
T'=T, \quad G'=G, \quad \bar G'=\bar G, \quad J'=J, \quad
H'=H-2i\b H, 
\nonumber\\[2pt]
&&
\bar H'=\bar H +2i\b \bar H, \quad
J^{(1)'}=J^{(1)}-2i\b J^{(1)}, \quad J^{(2)'}=J^{(2)}+2i\b J^{(2)},
\eea
and
\bea\label{automor3}
&&
T'=T, \quad G'=G+i\l \bar H +i\l H, \quad \bar G'=\bar G-i\l \bar H -i\l H,
\nonumber\\[2pt]
&&
H'=H+i\l \bar G -i\l G, \quad \bar H'=\bar H+i\l \bar G -i\l G,
\quad
J'=J-i\l J^{(2)} -i\l J^{(1)},
\nonumber\\[2pt]
&&
J^{(1)'}=J^{(1)}+2i\l J, \quad J^{(2)'}=J^{(2)}+2i\l J.
\eea
It is straightforward, although a bit tedious, to verify that 
Eqs.~(\ref{automor1})--(\ref{automor3}) do provide an automorphism of 
the $N=4$ SCA.

Since in two spatial and two temporal dimensions the Lorentz
group factorizes as $SO(2,2)=SU(1,1)\times SU(1,1)'$ one reveals
a tempting possibility to restore the full Lorentz symmetry for 
the $N=2$ string treated in the enlarged $N=4$ topological formalism. 
At the classical level this has been achieved~\cite{bg,bg1} by means of
enlarging the world sheet $d=2$, $N=2$ conformal supergravity multiplet
to the $d=2$, $N=4$ conformal supergravity multiplet.
The extra fields
transform nontrivially under ${SU(1,1)}_{outer}$ and render the full
action $SO(2,2)$ invariant. At the same time, as classically
the constraints $H=0$, $\bar H=0$, $J^{(1)}=0$, $J^{(2)}=0$
prove to be functionally dependent on those forming the $N=2$ SCA~\cite{ws,bg}
one still has the classical equivalence between the two formalisms. 
In the next section we turn to discuss 
how far one can get in restoring the Lorentz invariance 
at the quantum level.
 
\vspace{0.5cm}
\noindent

{\bf 3. Vertex operators}

\vspace{0.4cm}
Our main concern in this section will be the structure of tree amplitudes.
In particular, we shall study the constraints imposed on physical 
vertices by causality and cyclic symmetry of the amplitudes.
To this end we decompose
in modes (as we mentioned above one can choose the  NS representation
due to the spectral flow; in the relations below $n$ is an integer and 
$r$ is a half integer)
\bea\label{n2modes}
&&
L_n={\textstyle{\frac{1}{2\pi i}}} \oint dz~z^{n+1}~:T(z):, \quad
G_r={\textstyle{\frac{1}{2\pi i}}} \oint dz~z^{r+1/2}~:G(z):,
\nonumber\\[2pt]
&&
{\bar G}_r={\textstyle{\frac{1}{2\pi i}}} \oint dz~z^{r+1/2}~:\bar G(z):,
\quad 
J_n={\textstyle{\frac{1}{2\pi i}}} \oint dz~z^n~:J(z):,
\nonumber\\[2pt]
&&
H_r={\textstyle{\frac{1}{2\pi i}}} \oint dz~z^{r+1/2}~:H(z):,
\quad
{\bar H}_r={\textstyle{\frac{1}{2\pi i}}} \oint dz~z^{r+1/2}~:\bar H(z):,
\nonumber\\[2pt]
&&
J^{(1)}_n={\textstyle{\frac{1}{2\pi i}}} \oint dz~z^n~:J^{(1)}(z):,
\quad
J^{(2)}_n={\textstyle{\frac{1}{2\pi i}}} \oint dz~z^n~:J^{(2)}(z):,
\eea 
check the unitarity of the representation
\bea
{L_n}^{+}=L_{-n}, \quad {J_n}^{+}=J_{-n},
\quad {G_r}^{+}={\bar G}_{-r},
\quad
{H_r}^{+}={\bar H}_{-r}, \quad {J^{(1)}_n}^{+}=-J^{(2)}_{-n},
\eea
and work out the full algebra
\bea\label{n4alg}
&&
[ L_n, L_m ]=(n-m) L_{n+m} +{\textstyle{\frac {6}{12}}} (n+1)n(n-1) \d_{n+m,0},
\nonumber\\[2pt]
&&
[ L_n, G_r ]=({\textstyle{\frac 12}} n -r) G_{n+r}, \quad
[ L_n, {\bar G}_r ]=({\textstyle{\frac 12}} n-r) {\bar G}_{n+r},
\nonumber\\[2pt]
&&
[ L_n, J_m ]=-m J_{n+m}, \quad [ G_r, J_n ]=G_{n+r},
\nonumber\\[2pt]
&&
\{ G_r,{\bar G}_q \} =L_{r+q}-{\textstyle{\frac 12}}(r-q) J_{r+q}+
(r+{\textstyle{\frac{1}{2}}})(r-{\textstyle{\frac 12 }}) \d_{r+q,0},
\nonumber\\[2pt]
&&
[ {\bar G}_r, J_n ]=-{\bar G}_{n+r}, \quad
[ J_n,J_m ]=2n\d_{n+m,0},
\nonumber\\[2pt]
&&
[L_n, H_r]=({\textstyle{\frac 12}} n -r) H_{n+r}, \quad 
[L_n, {\bar H}_r]=({\textstyle{\frac 12}} n -r) {\bar H}_{n+r},
\nonumber\\[2pt]
&&
[ L_n, J^{(1)}_m ]=-m J^{(1)}_{n+m}, \quad
[ L_n, J^{(2)}_m ]=-m J^{(2)}_{n+m},
\nonumber\\[2pt]
&&
\{ G_r, {\bar H}_q \}={\textstyle{\frac 12}}(q -r) J^{(2)}_{r+q}, \quad
\{ {\bar G}_r, H_q \}={\textstyle{\frac 12}}(q -r) J^{(1)}_{r+q},
\nonumber\\[2pt]
&&
[G_r, J^{(1)}_n]=-2H_{r+n}, \quad [{\bar G}_r, J^{(2)}_n]=-2{\bar H}_{r+n},
\nonumber\\[2pt]
&&
[J_n, H_r]=H_{r+n}, \quad [J_n, {\bar H}_r]=-{\bar H}_{r+n},
\nonumber\\[2pt]
&&
[J_n, J^{(1)}_m]=2J^{(1)}_{n+m}, \quad [J_n, J^{(2)}_m]=-2J^{(2)}_{n+m},
\nonumber\\[2pt]
&&
\{H_r,{\bar H}_q \} =-L_{r+q}+{\textstyle{\frac 12}}(q -r) J_{r+q}
-(r+{\textstyle{\frac 12}})(r-{\textstyle{\frac 12}}) \d_{r+q,0},
\nonumber\\[2pt]
&&
[H_r, J^{(2)}_n]=-2G_{r+n}, \quad [{\bar H}_r, J^{(1)}_n]=-2{\bar G}_{r+n},
\nonumber\\[2pt]
&&
[J^{(1)}_n,J^{(2)}_m]=4J_{n+m}+4n\d_{n+m,0},
\eea
with other (anti) commutators vanishing. With respect to the $N=2$ algebra
physical states are defined to obey the standard relations
\bea\label{pstates}
&&
L_n~|\mbox{phys}\rangle=G_r~|\mbox{phys}\rangle=
{\bar G}_r~|\mbox{phys}\rangle=J_n~|\mbox{phys}\rangle=0 \qquad
n,r>0
\nonumber\\[2pt]
&&
L_0~|\mbox{phys}\rangle=J_0~|\mbox{phys}\rangle=0, 
\eea
where the absence of the normal ordering constants in the last two
relations follows from a rigorous BRST analysis\footnote{
In the $N=4$ framework the vanishing of the normal ordering constant 
for the operator $J_0$ is also required by the last commutator 
in Eq.~(\ref{n4alg}).}~\cite{bilal}. A simple inspection of the full $N=4$
algebra shows then that the modes $H_r$, ${\bar H}_r$, $J^{(1)}_n$, 
$J^{(2)}_n$ for $r,n>0$ automatically annihilate the physical 
states~(\ref{pstates}) if so do the zero modes of $J^{(1)}$ and $J^{(2)}$.
Thus the only extra conditions coming with the new generators are
\be\label{pstate1}
J^{(1)}_0~|\mbox{phys}\rangle=0, \quad J^{(2)}_0~|\mbox{phys}\rangle=0. 
\ee
Notice that this seems to be more restrictive then the situation 
at the classical level, where the constraints
$H=0$, $\bar H=0$, $J^{(1)}=0$, $J^{(2)}=0$ prove to be functionally dependent
and do not imply any new information as compared to that encoded into
the $N=2$ superconformal currents. The analysis of the 
relations~(\ref{pstates}) is well known~\cite{mukhi} which
has also been confirmed by the calculation of the one--loop partition 
function~\cite{mukhi,ov}. The spectrum of physical states consists of 
a single ground state, this being a massless scalar. As the operators
$J^{(1)}_0$ and $J^{(2)}_0$ give zero when acting on 
that state, the $N=2$ and $N=4$ spectra coincide.  

We now turn to discuss vertex operators. As the critical intercept equals 
zero ($L_0=0$ on physical states), a vertex operator 
valid for describing an asymptotic physical state 
must carry conformal spin zero. 
On account of Eqs.~(\ref{pstates}),(\ref{pstate1}) the natural guess is
\be
V_0(k,0)=:e^{ik {\bar x} +i{\bar k} x}:~,
\qquad k\bar k=0.
\ee 
Worth noting is that the vertex holds invariant with respect
to both the conventional $SU(1,1)$ and ${SU(1,1)}_{outer}$
(see the first line in the table above), 
thus exhibiting {\it the full} Lorentz invariance.
To construct a vertex operator $V(k,\bar k, z,\bar z)$ which would be 
capable of describing the emission of the bosonic state it is 
instructive to analyse the causality of the corresponding amplitudes. 
This implies, in particular, that spurious physical states 
decouple from physical processes  
\be\label{cause}
\langle \mbox{spur}|V(k,1)| \mbox{phys} \rangle =0,
\ee
where for simplicity we restricted ourselves to the three point 
function which proves to be sufficient for our purposes.
As is well known, the decoupling of spurious physical states of 
the form $\sum_{n>0} \langle \chi_n |~L_n$ with
\be\label{spur}
\sum_{n>0} \langle \chi_n |~L_n L_0=0 \rightarrow 
\langle \chi_n |~(n+L_0)=0,
\ee
requires $V$ to carry conformal spin 1
\be\label{physvertex1}
T(z)~V(z')\sim \frac{V(z')}{{(z-z')}^2}+\frac{\partial V(z')}{z-z'}.
\ee
Actually, from Eqs.~(\ref{physvertex1}) and~(\ref{n2modes}) one finds
\be
(L_n-n-L_0)V(k,1)=V(k,1)(L_n-L_0),
\ee
which when combined with Eq.~(\ref{spur}) renders the
amplitude~(\ref{cause}) vanishing. Given the matter fields,
the most general ansatz for the boson emission vertex is
\bea\label{verans}
&&
V(k,z,\bar z)=:\{\a \bar k \partial x +\b k\partial {\bar x}
+\l(\bar k \p)(k \bar\p)
+\m \underline{k \partial x} +
\g \underline{\bar k \partial {\bar x}}
+\rho(\bar k \p)(\underline{\bar k \bar\p})
\nonumber\\[2pt]
&& \qquad \quad
+\o(\underline{k\p}) (k\bar\p)\}e^{ik {\bar x} +i{\bar k} x}:~, \qquad
k\bar k=0,
\eea
with $\a,\b,\l,\m,\g,\r,\o$ some complex constants to be determined
below (here $\a,\b,\l,\o$ not to be confused with the infinitesimal 
parameters involved in the global ${SU(1,1)}_{outer}$ automorphism we 
discussed in the previous
section). The term like $(\underline{k\p}) 
(\underline{\bar k {\bar\p}})$ one could try to include into 
the ansatz above
proves to be redundant due to the identity
\be\label{idd1}
\e_{cp}\e_{\bar s \bar k}=-\eta_{c\bar s} \eta_{p\bar k} +
\eta_{c\bar k} \eta_{p \bar s}.
\ee
At this stage we refrain from requiring the vertex to respect the
full Lorentz symmetry, while by the very construction the ansatz does 
respect the conventional $SU(1,1)$.

Consider further the spurious physical states
generated by the fermionic currents ${\sum}_{r>0} \langle \chi_r |~G_r$ with
\be\label{spur1}
\sum_{r>0} \langle \chi_r |~G_r L_0=0 \rightarrow 
\langle \chi_r |~(r+L_0)=0.
\ee 
Taking into account the explicit form of $G(z)$ and the fact that $V$ has
conformal spin 1, one infers that the corresponding OPE can involve 
poles no higher than of the second order 
\be
G(z)~V(z')\sim \frac{U_1(z')}{{(z-z')}^2}+\frac{U_2(z')}{z-z'},
\ee
with $U_1(z)$ and $U_2(z)$ to be determined below. The 
amplitude~(\ref{cause}) then acquires the form
\be\label{spur3}
\sum_{r>0} \langle \chi_r |U_1(1)(r+{\textstyle{\frac 12}})+U_2(1)|
\mbox{phys} \rangle.
\ee
Assuming further that $U_1$ carries conformal spin $h$
\be\label{spinU_1}
T(z)~U_1(z')\sim \frac{hU_1(z')}{{(z-z')}^2}+\frac{\partial U_1(z')}{z-z'},
\ee
and taking into account Eq.~(\ref{spur1}) one immediately reveals 
that Eq.~(\ref{spur3}) vanishes provided
\be
h={\textstyle{\frac 12}}, \quad \mbox{and} \quad U_2=\partial U_1.
\ee
Thus the OPE of the fermionic current $G(z)$ with 
the boson emission vertex must be of the form
\be\label{opegv}
G(z)~V(z')\sim \frac{U_1(z')}{{(z-z')}^2}+\frac{\partial U_1(z')}{z-z'},
\ee
with $U_1$ having conformal spin $1/2$.

The current $\bar G(z)$ can be treated in the same way yielding the OPE
\be\label{opegv1}
\bar G(z)~V(z')\sim \frac{{\tilde U}_1(z')}{{(z-z')}^2}+
\frac{\partial {\tilde U}_1(z')}{z-z'},
\ee
with ${\tilde U}_1$ being a conformal field of spin $1/2$. 

So far we have considered spurious (bra) states generated 
by $G_r$, ${\bar G}_r$ and 
$L_n$ independently. It is instructive then to recall the 
anticommutator in the algebra $\{G_r,{\bar G}_r \}=L_{2r}$, $r>0$,
which relates the states and provides further information on the structure of
$V$. Actually, starting with the spurious state of the form
$\sum_{r>0} \langle \chi_{2r} |~L_{2r}$ with $\langle \chi_{2r} |(2r+L_0)=0$, 
one readily finds the identity
\bea\label{iden}
&&
\sum_{r>0} \langle \chi_{2r} |(2r+1)V(1)+\partial V(1) |\mbox{phys} \rangle=
\sum_{r>0} \langle \chi_{2r} |G_r\{ (r+{\textstyle{\frac 12}})
{\tilde U}_1 (1)+\partial {\tilde U}_1 (1)\}|\mbox{phys} \rangle
\nonumber\\[2pt]
&& \quad 
+\sum_{r>0} \langle \chi_{2r} |{\bar G}_r \{ (r+{\textstyle{\frac 12}})
U_1 (1)+\partial U_1 (1) \}|\mbox{phys} \rangle. 
\eea
Now we have to guess the OPE of $G(z)$ with ${\tilde U}_1(z')$
and $\bar G (z)$ with $U_1 (z')$. Since these have to match the left hand 
side of the identity~(\ref{iden}) the poles of the second order or higher are
not allowed (this is also prompted by the conformal spin which carry the 
fields $U_1 (z)$ and ${\tilde U}_1(z)$) and one finally has to set
\be\label{opegv2}
G(z)~{\tilde U}_1(z')\sim \frac{ {\tilde W}(z')}{z-z'}, \quad
\bar G(z)~U_1(z')\sim \frac{ W(z')}{z-z'}.
\ee
Substitutions of this in Eq.~(\ref{iden}) yields
\be\label{opegv3}
V=W+\tilde W,
\ee 
which together with Eqs.~(\ref{opegv}),(\ref{opegv1}),
(\ref{opegv2}) gives a closed set of equations to fix $V$. 
Finally, given the explicit form of the ansatz~(\ref{verans})
it is easy to verify the OPE's
\be
J(z)~V(z') \sim 0, \quad J^{(1,2)} ~V(z') \sim 0,
\ee
which hold independently of the value of the constant coefficients
entering the ansatz. This seems reasonable since, for instance, the
$U(1)$ charge is then conserved in the amplitude. 
The relations above imply also the decoupling of the spurious physical (bra)
states generated by $J_n,J^{(1,2)}_n$ with $n>0$. In a similar way,
making use of the algebra $[G_r, J^{(1)}_0]=-2H_r$, 
$[{\bar G}_r,J^{(2)}_0]=-2{\bar H}_r$, $[L_0, J^{(1,2)}_0]=0$
one can show the decoupling of those generated by $H_r$, ${\bar H}_r$
with $r>0$. Thus the decoupling of spurious physical states generated
by the currents extending the $N=2$ SCA to the $N=4$ SCA proves to be 
automatic and do not impose further restrictions on the form of the 
emission vertex.

In the $N=2$ framework the $U(1,1)$ invariance of the formalism
selects only the first three terms in the ansatz~(\ref{verans}).
Then a simple inspection of Eqs.~(\ref{opegv}),(\ref{opegv1}),
(\ref{opegv2}),(\ref{opegv3}) shows
that the only essential restriction coming along with the 
causality conditions reads
\be
\l=i(\b-\a).
\ee
This leaves two unspecified constants in the $U(1,1)$ invariant ansatz.
Calculating further the three point tree amplitude with physical {\it in} 
and {\it out} states
\be
A^{tree}_{right} (1,2,3)= \langle V_0 (k_1,\infty)V(k_2,1)V_0(k_3,0)
\rangle =-i (\a {\bar k}_2 k_3 + \b k_2 {\bar k}_3),
\ee 
one finds this to possess cyclic symmetry only when
\be\label{ab}
\b=-\a.
\ee
Up to an irrelevant number coefficient this reproduces the vertex found by 
Ooguri and Vafa~\cite{ov} within the alternative superfield approach
\be\label{ovvertex}
V(k,z,\bar z)=:\{ik\partial {\bar x} -i \bar k \partial x 
-2(\bar k \p)(k\bar\p) \}e^{ik {\bar x} +i{\bar k} x}:~, 
\qquad
k\bar k=0.
\ee
Given the vertex, tree amplitudes involving more than three legs 
prove to vanish~\cite{ov} which persists also 
to higher orders in perturbation expansion~\cite{berkvafa}.

Turning to the $N=4$ formalism, one is to consider the full 
ansatz~(\ref{verans}). The causality arguments now enforce
the conditions
\be
\l=i(\b-\a), \quad \rho=i\g, \quad \o=-i\m,
\ee 
where one has to use the identity
\be
\underline{k \p} (\bar k {\partial x}) -\underline{k {\partial x}} 
(\bar k \p)=
-(k\bar k) \underline{\p {\partial x}},
\ee
which is a consequence of Eqs.~(\ref{idd}) and~(\ref{idd1}). The new
terms entering the ansatz bring the extra contribution to the three
point function
\be
-i(\a {\bar k}_2 k_3 -\a k_2 {\bar k}_3 +\m \underline{k_2 k_3}
+\g \underline{{\bar k}_2 {\bar k}_3}),
\ee
and the last two terms reveal cyclic symmetry without any restrictions on
the coefficients $\g,\m$. Then a simple inspection of the table displaying 
the action of the ${SU(1,1)}_{outer}$ group on the elementary 
field combinations
shows that the correlator above {\it is not} ${SU(1,1)}_{outer}$ invariant
at any value of $\a,\m,\g$. A way out could be to allow the
constants to transform nontrivially under ${SU(1,1)}_{outer}$.
This type of reasoning has been advocated in Ref.~\cite{olaf} where two 
of the constants were identified with the gravitational and Maxwell 
string couplings. In our opinion, however, this does not fit well
neither the field theory framework nor the string theory framework.

Thus, we conclude that no Lorentz invariant boson emission vertex 
can be constructed within the $N=4$ topological formalism. The maximal
subgroup one can retain proves to be $U(1,1)$ which brings one back to the
Ooguri--Vafa vertex~(\ref{ovvertex}). Ultimately, 
our conclusion based on the old covariant quantisation proves to be in line
with the analysis of Refs.~\cite{berkvafa,ov3,bvw} which relied upon
the topological prescription for calculating the scattering amplitudes.

\vspace{0.5cm}
\noindent

{\bf 4. Concluding remarks}

\vspace{0.4cm}
To summarize, in this paper we reconsidered the issue of restoring the
Lorentz invariance in the quantum $N=2$ string. We used the passage to 
the equivalent $N=4$ topological formalism which raises the global 
automorphism group of the underlying superconformal algebra to the 
full Lorentz group. Being efficient in restoring the invariance at the
classical level, the $N=4$ formalism, however, is not powerful enough 
to do so in the quantum theory. In particular, the causality and cyclic 
symmetry of tree amplitudes prevent one from the
construction of a Lorentz invariant boson emission vertex. Although
asymptotic states still can be described by the Lorentz invariant vertices.

An interesting continuation of this work could be a rigorous BRST 
analysis, although in the BRST framework we do not see means which 
could impact the conclusion drawn in this paper.

\vspace{0.3cm}

\noindent
{\bf Acknowledgements}\\[-4pt]

\noindent
A.G. thanks E.A. Ivanov for helpful comments.
This work was supported by INTAS grant No 00-0254 and by the Iniziativa 
Specifica MI12 of the Commissione IV of INFN. 


\end{document}